\documentclass[final,3p]{elsarticle}

\usepackage{lineno}
\usepackage[colorlinks=true,breaklinks=true,pdftex]{hyperref}
\usepackage{amsmath}

\usepackage{soul}
\usepackage{wrapfig}
\usepackage{xcolor}
\modulolinenumbers[1]

\journal{Computer Aided and Geometric Design}

\biboptions{authoryear}

\newcommand{\dataset}{Real-World Textured Things}
\newcommand{\ds}{RWTT}
\newcommand{\texmetro}{TexMetro}

\newcommand{\rev}	[1]{{#1}}

\begin{document}

\begin{frontmatter}

\title{\emph{\dataset{}}: a Repository of Textured Models Generated with Modern Photo-Reconstruction Tools}

\author[unimi]{Andrea Maggiordomo\corref{cor}}
\ead{andrea.maggiordomo@unimi.it}
\author[isti]{Federico Ponchio}
\ead{federico.ponchio@isti.cnr.it}
\author[isti]{Paolo Cignoni}
\ead{paolo.cignoni@isti.cnr.it}
\author[unimi]{Marco Tarini}
\ead{marco.tarini@unimi.it}
\cortext[cor]{Corresponding author}
\address[unimi]{University of Milan, Italy}
\address[isti]{ISTI - CNR, Italy}

\begin{abstract}
We are witnessing a proliferation of textured 3D models captured from the real world with automatic photo-reconstruction tools by people and professionals without a proper technical background in computer graphics.
Digital 3D models of this class come with a unique set of characteristics and defects -- especially concerning their parametrization -- setting them starkly apart from 3D models originating from other, more traditional, sources.
We study this class of 3D models by collecting a significant number of representatives and quantitatively evaluating their quality according to several metrics. These include a new \emph{invariant} metric we carefully design to assess the amount of fragmentation of the UV map, which is one of the main weaknesses potentially hindering the usability of these models. 
Our results back the widely shared notion that models of this new class are still not fit for direct use in downstream applications (such as videogames), and require challenging processing steps.
Regrettably, existing automatic geometry processing tools are not always up to the task: for example, we verify that available tools for UV optimization often fail due mesh inconsistencies, geometric and topological noise, excessive resolution, or other factors; moreover, even when an output is produced, it is rarely a significant improvement over the input (according to the aforementioned measures).
Therefore, we argue that further advancements are required by the Computer Graphics and Geometry Processing research communities specifically targeted at this class of models. Towards this goal, we share the models we collected in this study in the form of a new public repository, \emph{\dataset} (\ds), intended as a  benchmark to systematic field-test and compare future algorithms.
\ds{} consists of 568 carefully selected textured 3D models representative of all the main modern off-the-shelf photo-reconstruction tools. The repository is web-browsable by the metadata we collected during our experiments, and comes with a tool, \texmetro{},  providing the same set of measures for generic UV mapped datasets.
\end{abstract}

\begin{keyword}
Benchmarks \sep
Photo-reconstruction \sep
3D Acquisition \sep
Quality Measures \sep
Surface Parametrization \sep
UV maps \sep
Textures \sep
Real-World Models 
\end{keyword}

\end{frontmatter}

\section{Introduction}\label{sec:intro}

In the last decade, many research and implementation efforts have culminated in the development of algorithms, technologies, and suites that facilitate creating and publishing 3D content by professionals and hobbyists, not necessarily equipped with a technical background in computer graphics or geometry processing.
Among the most influential and exciting advancements, photo-reconstruction techniques played a primary role, impacting a variety of application fields such as artwork and architectural documentation, environmental monitoring, data acquisition for manufacturing, 3D assets creation.
Commercial and free software solutions are now available to both professionals and hobbyists, and used to generate accurate 3D digital copies of real-world objects in an almost fully automatic way.

This democratization of 3D content creation determined the diffusion of textured 3D digital models sharing certain specific characteristics, differing both qualitatively and quantitatively from those of 3D representations typically targeted -- and assumed -- by the large majority of academic works on Computer Graphics and Geometry Processing.
Consequently, there is now an increased demand for robust techniques designed to process, manipulate, and ameliorate this class of 3D model representation, and that further research efforts are required to fill this gap.

\paragraph{Shared Characteristics of typical Real-World, Textured models}

Digital 3D models acquired with photo-reconstruction techniques share a combination of traits that are not found in any other class of 3D models (such as 3D meshes digitally sculpted or manually designed by professional 3D modelers, range-scanned models, or procedurally generated models).

The \textbf{meshing} is typically high-resolution but low quality, triangle-based, and completely irregular.
The frequent inaccuracies in the 3D reconstruction result in geometrical noise and meshing defects. \rev{While many solutions have been proposed to face these problems, (like for example \cite{Attene2013,ju2004robust,chu2019repairing}) the proposed approaches focus on the geometrical aspect and do not take into account the presence of an existing parametrization.} Even more importantly,
geometry is but one aspect of the description: photo-reconstructed meshes are enriched with one or multiple high-resolution textures derived from the same set of photographs used to reconstruct the shape; these are a central part of the description of the object, as well as the object UV map, i.e. the parametrization defined to map the texture data onto the object surface.

\begin{wrapfigure}{R}{0.46\textwidth}
\centering
\includegraphics[width=0.42\textwidth]{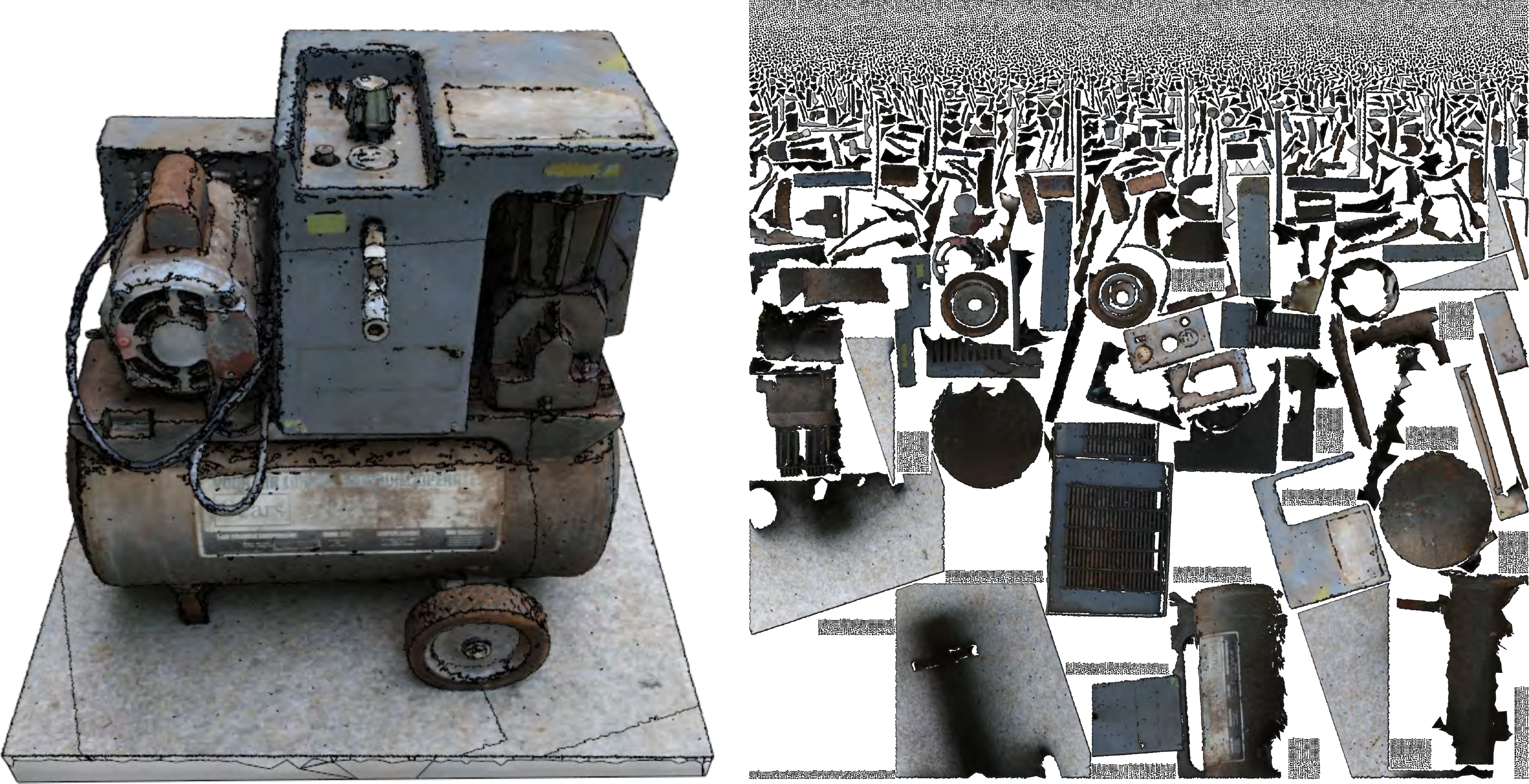}
\caption{One model of the RWTT dataset, with its texture and UV map.}
\label{fig:intro:atlas}
\end{wrapfigure}

The \textbf{UV maps} of the 3D models present the most distinctive traits, directly reflecting the way they have been constructed (see Sec.~\ref{photoscan} below).
Specifically, they are atlas-based parametrizations where the atlases are typically extremely fragmented and with irregularly shaped charts, to a much larger extent than other categories of models (see for example Fig.~\ref{fig:intro:atlas}).

This exacerbates by several times over all the numerous defects \rev{and costs} associated to texture seams of the UV maps,
\rev{impacting any downstream application.}
The reader is referred to \cite{tarinirtm} for a more comprehensive analysis of these issues.
\rev{In summary, they include poor GPU memory usage, due to a combination of packing inefficiency and the
need to replicate texels on a small band around each seam to avoid noticeable texture ``bleeding'' artifacts;
unavoidable minor rendering artifacts, appearing to are due to the inconsistent bilinear sampling on the seams (even in the presence
of said texel duplication); inability to use MIP-mapping without incurring in additional bleeding artifacts (unless the width of the texel duplication bands is enlarged exponentially to the maximal MIP-map level); vertex duplications in the data structure encoding the mesh, which is necessary to be able to encode the seams in the first place (implying a small additional cost both in memory and transform workload); consequent constraints on processing or preprocessing performed over the mesh data structure, for example hindering to some extent compression, streaming coarsening, construction of multi-resolution structures (such as \cite{nexus05}). The latter factors are due to the necessity of the texture seams to be represented by (duplicated) edges in the meshing.}

\paragraph{Contributions}

With this in mind, in this paper we offer the following contributions.
We present an \textbf{analysis} of the aforementioned class of 3D models and their characterizing features, by collecting a large sample of representative models and performing systematic measurements according to several criteria we identify as relevant to geometry processing algorithms and computer graphics applications in general, focusing our analysis on the UV maps as its presence is the defining characteristics of these models.

To support our analysis, we design a \textbf{new quality measure}, the atlas \emph{crumbliness}, that captures the severity of texture seams in a given UV map.
This new measure is compact, easy to compute, and has desirable properties such as being invariant to UV scale and mesh subdivision.

Since our analysis shows that the quality of the parametrizations in our sample data is low, we \textbf{evaluate} commercial, open-source and academic software on the task of recomputing UV maps for the 3D models, and show how existing algorithms and implementations typically either fail, are unable to improve the quality of these maps or are not practical.

We make our \textbf{dataset}, Real-World Textured Things (RWTT), publicly available on the web.
The dataset has been curated and organized to facilitate batch-processing, and is suited to test, benchmark, and challenge geometry processing algorithms, both existing and future.
Overall, \ds{} features 568 models generated by real-world users of the aforementioned photo-reconstruction technologies and released with permissive licenses, striving to represent the wide variety of user categories, used photo-reconstruction tools (including all existing off-the-shelf ones), application fields and styles.
Our aim is to provide a standard, realistic, and challenging testbed for algorithms, able to indicate their usability in the real world.
All models in RWTT come pre-assessed according to the measures we collected, and are browsable and searchable by them.

Finally, we provide the \textbf{measuring tool} we used to evaluate the models, as a publicly available, stand-alone software application: \texmetro{} (Sec.~\ref{sec:texmetro}).
\texmetro{} automatically grades user-provided datasets (in the same way dataset of RWTT are graded), and is intended as a tool to help researchers and practitioners assessing and comparing the effectiveness of future algorithms and techniques
designed to target this class of models. In this, we are inspired by the impact that tools of this kind had in the geometry processing community (Sec.~\ref{sec:related2}).

\subsection{Background in 3D Photo-Reconstruction}
\label{photoscan}

Photo-reconstruction techniques (\cite{hartley2003multiple, Seitz:2006, remondino2014state}) generate a textured 3D mesh starting from a set of high-resolution 2D pictures shot at the subject.

There are many possible variations, but the general principle can be summarized as follows.
First, Structure-From-Motion techniques are used to estimate camera parameters: feature points are identified on each input image, then matching features points are identified across images using similarity measures. In a global adjustment phase, a set of camera positions is estimated, together with 3D positions of feature points, to maximize consistency, e.g.\ by minimize total back-projection errors or maximize color consistency across images.
Once the camera parameters have computed, Multi-View-Stereo algorithms are employed to construct a dense point cloud.
Finally, the point cloud is processed to remove noise and outliers, and fused into one polygonal mesh representation.
The mesh representation is used to estimate the occlusion and visibility of each pixel in the picture.
A subset of the visible pixels are then copied into one or more high-resolution textures, which provide the majority of the visual detail of the final result.
Textures are sometimes post-processed to alleviate shadow artifacts, color jumps when switching from one source image to another, or can optionally store additional material and surface information (e.g.\ surface normals).

Each step of this chain constitutes a challenging problem, which has been studied in separation or within complete systems.
The research on each sub-problem has recently matured enough so that a large number of off-the-shelf commercial tools integrating implementation of all the used techniques, both commercial and open-source, are now available \cite{realitycapture, autodesk, agisoft, zephyr, pix4d, openmvg, meshlab, contextcapture, itseez}.

For our purposes, the most relevant part is the construction of the UV map.
Parametrization construction techniques traditionally employed in other contexts are often not adequate to the task, in part because these methods often lack the full automatism required in this scenario.
Moreover, many methods (automatic or user-assisted) are hindered by the characteristics of the meshing (i.e.\ its large resolution and various defects).
At the same time, the presence of a set of images that features every part of the object surface (by construction) unlocks UV map construction strategies that \rev{are available only in this scenario}.
Namely, each mesh triangle can be assigned to one of the photos using heuristics (based on estimated visibility or other factors).
Thus defines islands inside the photos which then just need to be packed into (one or more) final textures.
Note also that that mapping distortions, traditionally of central importance in the effort of constructing a good quality UV map for a given mesh, are not particularly relevant in this context.

Differently from most other scenarios, where a UV map is constructed before the texture is filled with the final signal, here the texture content (color from images) \emph{precedes} the UV map construction. Because the source of the texture colors are limited to the images, there is no gain in diminishing the mapping distortions which can already present from these images to the 3D surface.
Therefore, instead of diminishing the distortion by optimizing a suitable map, the optimization strategy can only consist in choosing the least distorted input image to use as a source for a given triangle, which is (for example) the image offering the most orthogonal view of that portion of the surface (barring other considerations, such as occlusion, presence of highlights, optimization of island shapes, and others).

\rev{These factors determine that the UV found on this important class of models features unique and specific characteristics, motivating the present work.}

\section{Related Work}\label{sec:related}  

The practice of creating good, comprehensive, realistic, challenging benchmarks is well recognized, and there is no need to state their importance in promoting good practices and improving repeatability, comparability, and validation of research.
In this section we briefly summarize other dataset contributions made to the geometry processing community and public repositories of textured models. There are many geometric and computer vision datasets available that target a large variety of tasks.

Datasets for machine learning generally feature large collections of objects, and come with metadata to facilitate training and validation of learning-based algorithms. The ModelNet benchmark \citep{wu20153d} is a collection of more than a hundred thousands CAD models, designed to train and benchmark object recognition algorithms. ShapeNet \citep{shapenet2015} is a large scale dataset of annotated 3D models. Datasets designed to benchmark recognition and classification algorithms provide metadata about categories and annotations to facilitate the training of the geometric networks. On a similar note, the ABC dataset \citep{abc} is a massive dataset of one million CAD models proposed to test the effectiveness of learning-based approaches to the solution of common geometric problems such as normals computation.

Related to object scanning, there exists a large quantity of RGB-D datasets. The review of those datasets is beyond our scope, and we refer the reader to the excellent survey \cite{firman-cvprw-2016}.

Scanning campaigns have produced popular model repositories, such as the Digital Michelangelo Project \citep{levoy2000digital} and the Stanford Scanning Repository \citep{StanfordScanRep}, whose models are known to the entire computer graphics community.
None of these models, however, have textures, and they have been generated using 3D scanners; in contrast, the models featured in our dataset are created with photo-reconstruction algorithms.

The Princeton Shape Benchmark \citep{shilane2004princeton} is a large collection of low-resolution models with very simple geometries. The \cite{largegeom} dataset focuses instead on high-resolution models. Both datasets are now rather outdated.

More related datasets are the Thingi10K dataset \citep{thingi}, which features real-world objects and targets 3D scanning. \citep{Choi2016} is a large dataset of more than ten thousand object scans and the associated RGB-D data. The collection is extensive and, like ours, features object obtained from 3D reconstruction but again without textures.

SceneNN \citep{scenenn-3dv16} and ScanNet \citep{dai2017scannet} feature annotated scenes reconstructed from RGB-D video sequences, and among the existing datasets are, to our knowledge, the most similar to ours. However, they do not contain models reconstructed with photogrammetric techniques, features mostly indoor scenes and the 3D reconstructions are not particularly accurate.

As outlined above, geometric and computer vision datasets seldomly include textures. Textured models can be found in asset stores such as the Unity Asset store \citep{unity}, the Unreal Marketplace \citep{unreal}, and online repositories such as TurboSquid \citep{turbosquid}. Models from these sources however are typically non-free, sculpted, and textured by hand by 3D artists.

In contrast to all of the above, \dataset{} is a collection of large, textured 3D models generated with a variety of photo-reconstruction software tools, and features models published with permissive licenses to allow use in research and other practical contexts.

\subsection{Quality Measurement Tools} \label{sec:related2}

The idea of providing public tools intended to be used to measure specific quality measures of user-provided models has several predecessors.
Among the most notable, is \emph{Metro}, a tool presented in \cite{metro1998}, which computes and reports the Hausdorff distance
between pairs of surfaces provided as triangular meshes.
Over the last two full decades, \emph{Metro} has helped to assess and compare vast amounts of algorithms, e.g. for surface simplification, remeshing, surface reconstruction, and other geometry processing tasks -- as reported by thousands of research papers \citep{metro:scholar}.
As an additional benefit, the adoption of \emph{Metro} contributed to uniform the way quality measures are defined and
implemented. Overall, \emph{Metro} has proven to be a tremendous aid in the research community striving
to invent new algorithms and solutions toward a given set of geometry processing problems.

\section{Dataset}\label{sec:overview}

As mentioned in the introduction, one of the motivations behind the creation of the \dataset{} dataset is to provide researchers and developers a challenging testbed of textured 3D models indicative of the actual models that can be encountered in practice.
Rather than creating the models in-house, we have collected publicly available models created and released by professionals practitioners and hobbyists.
This strategy has been successfully employed by other authors (e.g. \cite{thingi, abc}) and has the important advantage of allowing us to include models generated with various software packages, which is crucial to obtain a representative sample of the 3D data that can be found in the real world.
In particular, it is reasonable to expect the emergence of some common traits characterizing the geometry and textures of the models in the dataset according to the implementation details and algorithmic choices of different reconstruction pipelines.

\begin{figure}
\centering
\includegraphics[width=\linewidth]{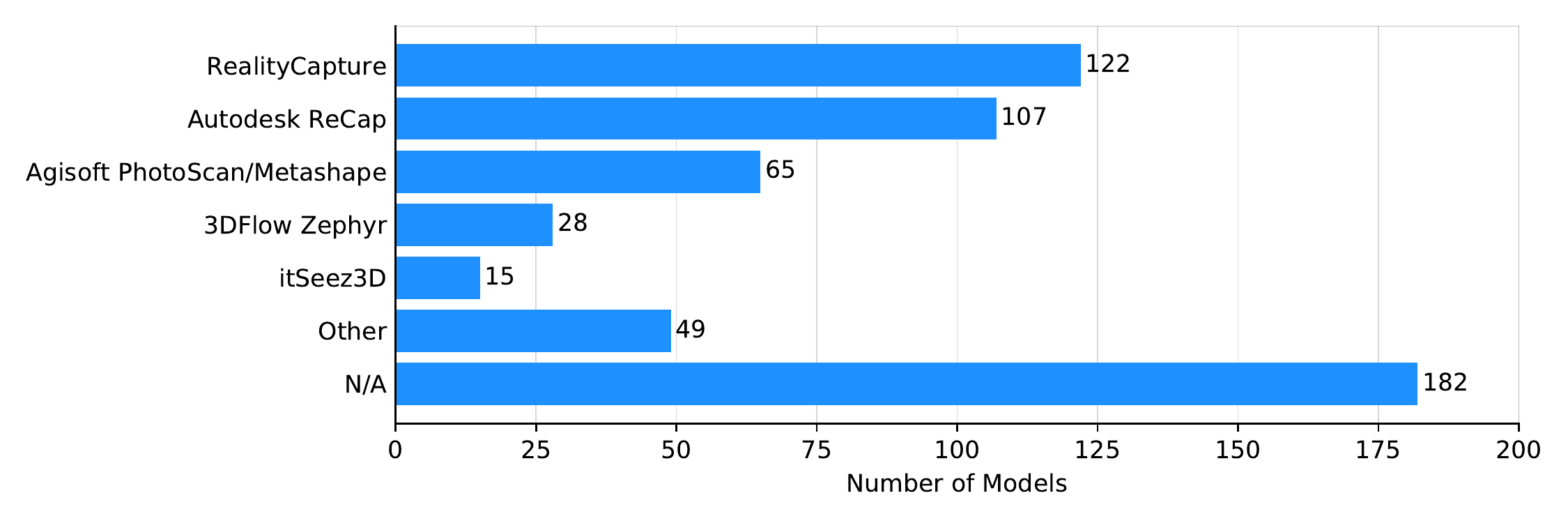}
\caption{Overview of the different software tools used to create the models featured in \dataset{}. There are at least 18 different reconstruction pipelines represented in the dataset; here for compactness we only report the most frequent occurrences.}
\label{fig:dataset:tools}
\end{figure}

At the time of writing, the dataset features 568 models created by 129 unique authors using at least 18 different acquisition pipelines.
Figure \ref{fig:dataset:tools} reports a taxonomy of the models featured in the \ds{} dataset categorized according to the reconstruction software used for their generation; this information was obtained from the textual descriptions and tags available when the models were retrieved.

\rev{We have collected models that were published across 6 years} from 2013 to 2019 from three sources: SketchFab \citep{sketchfab}, the Smithsonian 3D Program \citep{smithsonian} \rev{and the Zamani Project \citep{ruther2020}}.
SketchFab is a popular online repository for publishing and sharing 3D content used by thousands of users and constantly growing; it features a large collection of 3D models reconstructed from photographs and conveniently exposes a web API to automatically retrieve metadata such as the model author, description and tags among others.
The Smithsonian 3D Program is a digitization project from the Smithsonian Institution published on the web and made available a selection of 3D reconstructions of artifacts and relics.
\rev{The Zamani Project is an ongoing effort to document important heritage and archaeological sites in Africa by providing, among other material, high resolution textured 3D models}.

We have been careful to almost exclusively include models that have been released with permissive licenses to allow anyone using the dataset to freely share their results: all of the models from SketchFab are released under various declinations of the Creative Commons 4.0 license; the models from the Smithsonian 3D Project follow the ``Smithsonian Terms of Use'', allowing fair use as long as the original source is cited. We plan to keep the dataset updated by adding new models in the future.

\subsection{Dataset Curation}\label{sec:organization}

The dataset is organized as a collection of folders using a progressive naming scheme to facilitate the automatic processing of the 3D models.
Each model in the dataset is assigned a progressive identifier $ID$ and is stored in a folder named RWT$ID$.

For each model, we provide the following data:
\begin{itemize}
\item The 3D model in OBJ format and its material file
\item The textures referenced by the material file
\item A JSON file with the provenance metadata of the model
\item A JSON file with geometric and parametrization measures
\end{itemize}

Although the original models and textures came in a variety of formats, we have converted all of them to unique formats for convenience; in several cases, manual fixing of the object files was required to restore correct material libraries and texture image references.
All the models are provided as OBJ files, and the textures as PNG images.
Whenever necessary, the conversion only involved the file formats while the 3D and image data was left untouched; the models archived in their original formats are available separately.
The textures we provide generally encode surface colors; however, several models also come with high-resolution textures storing other signals such as surface normals, roughness, specularity, and ambient occlusion.

Each model comes with its \emph{metadata.json} file reporting information such as the name of the 3D model, the author, the license under which the model was released, a textual description of the model provided by the author, the date in which the model was published, a collection of tags chosen by the author, and a resource identifier.
For models obtained from SketchFab the resource identifier is the URI assigned from SketchFab itself, whereas for the ones obtained from the Smithsonian 3D Program the URI is the link to the web-page showcasing the model.
When available, we also include the software used to generate the model and a semantic category we assigned to the models for classification purposes.

Finally, we also include a \emph{texmetro.json} file generated with the \texmetro{} tool.
This JSON file contains information about the geometry, parameterization, and textures of the models; a detailed description of the \texmetro{} tool and its output is given in section \ref{sec:texmetro}.

\section{Quality Measures for Textured 3D Meshes} \label{sec:measures}

The quality of a textured mesh can be assessed by several quantifiable criteria, reflecting the resources that will be necessary to deploy the model (e.g.\ store it, stream it, or render it)
and the quality of the rendered images, and thus the degree of usability of a given model in a specific context.

Models in \ds{} are scored according to a suitable set of such measures, which is described in this Section.
\rev{Different measures can impact downstream applications in different ways. Collectively, the set of measures reveals
the peculiarity of the class of meshes we are focusing on, setting it apart from meshes commonly encountered in contexts
different from automated photo reconstruction.

Our set of measures includes: general criteria (such as data size, surface genus), UV map measures (concerning the parametrization),
and defects (number of occurrences of typical inconsistencies).}

\rev{\subsection{General Measures}}

Geometric and topological measures include the model resolution in terms of faces, vertices, and edges, the surface area,
the number and total length of the mesh boundaries and the surface genus. \rev{Topological information is relevant since many geometry processing algorithms require strict assumptions on the topology of the input data.}
Texture resolution and number of textures, are also meaningful measures, as these are all quantities that would make this data impractical to be handled by algorithms that do not scale well with the input size.

\subsection{UV map Measures}

\rev{The quality of the surface parametrizations and texture maps are the main focus of our analysis and we list here the metrics we have adopted.}

\subsubsection{UV Occupancy}

We define UV occupancy as the ratio of the number of texels strictly inside any UV triangle over the total number of texels.
A better UV occupancy results in a higher texel density over the surface for a fixed texture resolution.
or lower GPU memory footprint for the same rendering quality.
Occupancy is affected by the quality of the atlas packing: tightly packed atlases waste fewer texels between charts.
\rev{Atlases that are highly fragmented have also a worse occupancy.}

\subsubsection{Atlas Crumbliness and Solidity}

\begin{figure}
\centering
\includegraphics[width=\linewidth]{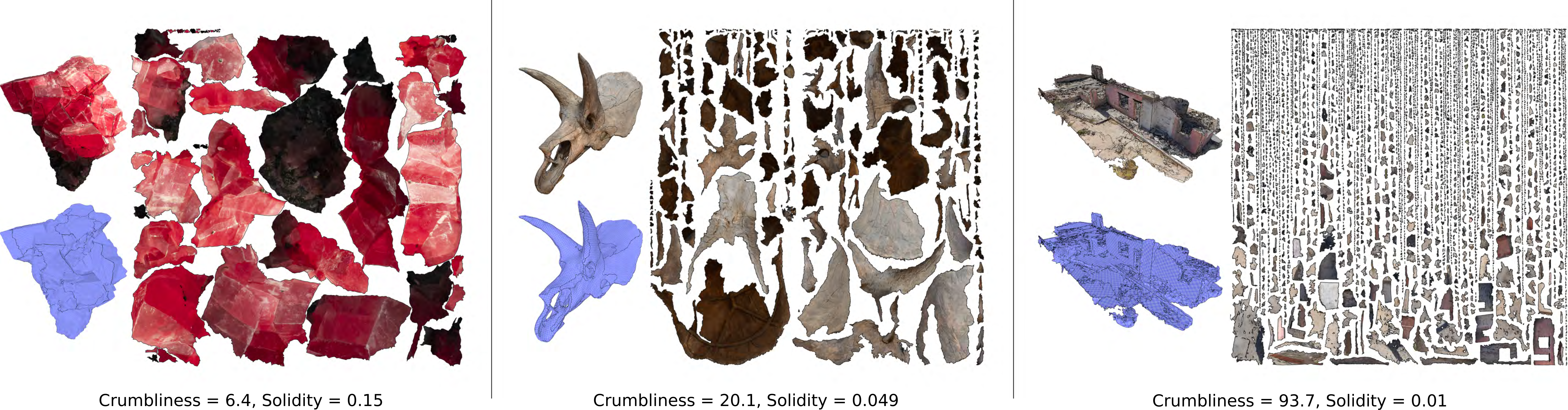}
\caption{Three models with their texture seams highlighted. While some texture maps featured in our dataset have a relatively low chart count (left), many others exhibit a surprisingly large number of texture seams unrelated to the geometrical and topological complexity of the models (center and right). Crumbliness and solidity values conveniently and compactly model the fragmentation of the texture atlas.}
\label{fig:seams}
\end{figure}

\rev{As summarized in the introduction, the presence of texture seams is somewhat detrimental to the usability of the dataset, for several different reasons. While texture seams are, to some extent, necessary in any textured mesh, we focus on the presence of redundant texture seams, due to the fragmentation of the UV atlas, and independently from the complexity of the 3D reconstructed shape (see figure \ref{fig:seams}). Therefore, we need to adopt a quality measure to assess the level of fragmentation of an existing texture atlas.}

\rev{Because there is no standard, we devise a new measure for this characteristic, which is one defining feature of the class of 3D models of our dataset. We considered several alternatives. We observe that the number of charts is not, alone, a good descriptor, for two reasons: first, charts can have varying sizes and boundaries of varying complexity; second, the chart may include any amount of ``non-disconnecting'' cuts, i.e.\ cuts that do not split the chart.
The number of seams edges, or the ratio of seam edges to non-seam edges, would also not be adequate: these measures would be affected by a regular subdivision of the mesh, which does not affect the level of fragmentation. Similarly, total length of seam edges would not be invariant to a rescaling of UV.}

Instead, we define a new measure, which we term \emph{crumbliness}.
Crumbliness is the ratio of the total length of the perimeter of the atlas charts (summed over all charts) over to the perimeter of a circle having the same area as the summed area of all charts.
Because the circle is the shape maximizing the area for a given perimeter, this measure has 1 as the lower (and best) bound. In formulas, if $a_i$ is the area of chart $i$ and $p_i$ is its perimeter, the crumbliness of the atlas $C$ if given by
\begin{equation*}
C = \left( \sum_i{ p_i }\right) \left({ 4 \pi \sum_i{ a_i } }\right)^{-\frac{1}{2}}
\end{equation*}
\rev{Crumbliness is independent of the mesh resolution, invariant under element subdivision,
and scale-independent. As such, it can be used to compare atlases of different models and for textures of varying sizes.

We define \emph{atlas solidity} as the inverse of crumbliness. Compatibly with many other quality measures,
atlas solidity has an upper bound of 1, and higher values indicate a better quality.

\paragraph{Uniformity of texture sampling} \label{sec:textures:uvscaling}

A well-understood characteristic of any parametrization (the UV map) is the distortion it introduces. In the ideal case, the mapping between the surface and the texture preserves areas, angles, or lengths (implying both areas and angles); the failure in preserving these features is measured by various \emph{distortion measures}. Conformal distortion reflects angles variation, area distortion reflects area variation, and isometric distortion reflects length variation. A multitude of different measures have been presented in the literature (for each case), usually to minimize them during automatic UV map construction (see \cite{floater2005surface} for a survey, and, for more recent examples, \cite{nollstricker,scaffold2017,poranne2017autocuts} among many others).

In our context, which is texture mapping, distortion is relevant because it directly determines how texture samples (texels) will be distributed on the surface. Area distortion translates into nonuniform sampling density, and conformal distortion translates into anisotropic sampling.

Sampling density can be hugely varying in models of \ds{},
due to the different relative resolution of the pictures used to reconstruct the models, and the different
distance between the acquiring camera and the captured object.

Instead of adopting any specific measure of area distortion, we choose to measure directly its effect, by analyzing the sampling density of texels over the mesh.
}

\begin{figure}
\centering
\includegraphics[width=\linewidth]{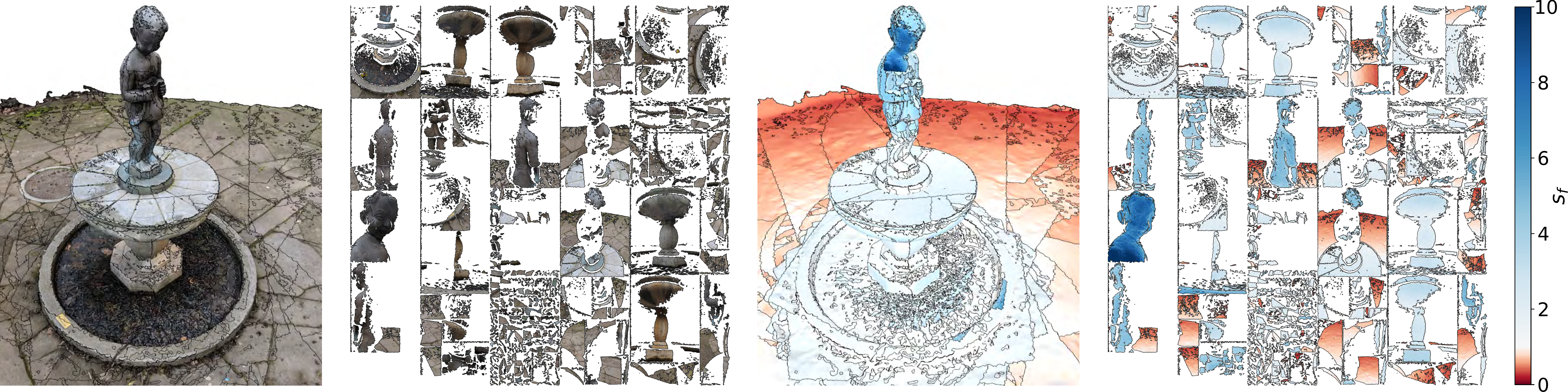}
\caption{Visualization of the scalar field $s_f$ on a model affected by uneven sampling frequency. Regions densely and sparsely sampled are colorized in blue and red respectively, reflecting the amount of color information encoded in the texture image. In this case, the model was clearly reconstructed using close-ups of the statue with the pavement left in the background.}
\label{fig:pfboy}
\end{figure}
\rev{
To this end, we define a per-element scalar quantity $s_f$ that represents the texture sampling resolution as the ratio between the local (per-element) and global (per-mesh) change of scale from 3D to UV:
\begin{equation*}
s_f = \frac{A_{\text{UV}}(f) / A_{\text{3D}}(f)}{\sum_f A_{\text{UV}}(f)/ \sum_f{A_{\text{3D}}(f)}} .
\end{equation*}
Where $A_{\text{UV}}$ and $A_{3D}$ indicate element areas computed in UV and 3D space respectively.
The local value of $s_f$ is close to 1 for uniform sampling distributions, and approaches zero as the mapping resolution of $f$ gets smaller relative to the global resolution of the map and becomes greater than one otherwise.
Therefore the scalar field $s_f$ indicates, over a model, where the color signal is undersampled and oversampled w.r.t. to the average sampling density of one given model.
Figure \ref{fig:pfboy} shows an example of how uneven sampling can impact the quality of a textured model when it is rendered.

As an aggregate measure for a model, we provide the variance of $s_f$ over the mesh (its average is 1.0 by construction).
Small values reflect more uniform sampling, and the minimal value 0 represents a perfectly uniform sampling. Note that this is
not necessarily desirable for models in our class; in fact, even a vastly nonuniform sampling density can correctly reflect the different semantic importance of certain parts of the model relative to others (such as an object in focus vs. background) or different quantities of available color data.

\subsubsection{UV Conformal Distortion}
\label{sec:textures:conformality}

Almost conformal mappings (i.e.\ UV maps exhibiting low conformal distortions) are desirable because they imply a more isotropic texel sampling distribution over the surface, which positively affects the quality of the signal reconstruction and avoids texture stretching artifacts.
Because this factor is not captured by the sampling density, we measure conformal distortion separately.

To this end, we adopt a formulation similar to \cite{circlepatterns}:
which defines the quasi-conformal distortion ($qcd$) of a map as the ratio of the smallest to largest singular value
of the Jacobian of the mapping. This is the inverse of what is proposed in \cite{circlepatterns}; we opt for this formulation so
that, the maximal of value 1.0 corresponds to the best case (a perfectly conformal map), consistently with most other quality measures in our set.
}

\rev{
\subsubsection{Discrepancy in the Signal Reconstruction at Seam Edges}

One specific defect linked to the presence of seams is the potential discrepancy of the texture signal on the two sides
of the texture seam, which can be due to either mismatching texel values, and to an inconsistent bilinear interpolation between texture values.
This does not depend on the UV map alone, but on the texture values as well.
Therefore, we include a separate measure to assess the extent of this problem for a given specific model.

To this end, we adopt the measure recently defined (for the purpose of minimizing it) in \cite{seamless}, which, for completeness, we describe in the following.

Given an UV edge $e$, $e(t)$ denotes the linear combination of its endpoints, with the parameter $t$ restricted to the $[0,1]$ interval.
A (manifold) edge $e$ of the 3D mesh is a seam edge if corresponds to two non coinciding UV edges $e_1$, $e_2$.
Let $f$ be the function returning the texture value for a given UV coordinates, using bilinear interpolation of the four closest texels. The reconstruction discrepancy at a seam edge $e$ is then defined as:
\begin{equation*}
D(e) = \int_{0}^{1} | f(e_1(t)) - f(e_2(t)) | \, dt .
\end{equation*}
As a global discrepancy measure, we report the average discrepancy of all the seam edges in a 3D model
\begin{equation*}
D(S) = \frac{\sum_{e \in S} \|e\| \cdot D(e)}{\sum_{e \in S} \| e \|}
\end{equation*}
where $S$ denotes the set of seam edges, and the each seam edge contribution is weighted by its length $\|e\|$ to capture the visibility and severity of the resulting artifacts independently from the mesh subdivision.
In our implementation, edge discrepancies are estimated using discrete samples.
}

\subsection{Defects}
\rev{
One defining characteristic of datasets of \ds{} is the common occurrence of defects and inconsistencies of specific nature. These issues are usually not considered when geometry processing algorithms are developed on paper, which often assume to be receiving sanitized input. Different issues can be encountered in the geometric part, or the parametrization part.

Defects of geometry include extremely irregular triangulations, zero-area triangles, topologically degenerate triangles (referring the same vertex more than once), and unreferenced or duplicated primitives.
Topological defects include errors such as local loss of two-manifoldness, holes, and incomplete meshes.
Another potential source of issues is the presence of ``topological noise'', i.e., small handles unrelated to the ideal topology of the reconstructed 3D shape (typically generated during the surface reconstruction phase from noisy dense point clouds), which is revealed by incongruous genus number with respect to the overall model shape.

Defects of the UV maps include loss of injectivity, resulting in the same texture data to be erroneously mapped into multiple parts of the mesh. Overlaps in the UV map can be either
\emph{local}, when triangles flip in UV space and change orientation relative to their neighbors, or \emph{global}, when the boundary of one or several charts intersect in UV space.
While they cause the same problem,
it is useful to distinguish between local and global overlaps, because the strategies available to counter them are typically distinct.
Additionally, several dataset feature only partial maps, that is UV maps that do not map a model surface in its entirety; this is likely due to insufficient photographic data.
}

\section{Measurements of RWTT Models: an Analysis}\label{sec:analysis}

We applied our set of measures overall models of RWTT. In this section, we offer an
analysis of the results, which serves as a picture of the merits and limitations
of the currently available off-the-shelf photo-reconstruction software.

\subsection{Measures on Geometry}\label{sec:geometry}

\begin{figure}
\centering
\includegraphics[width=\linewidth]{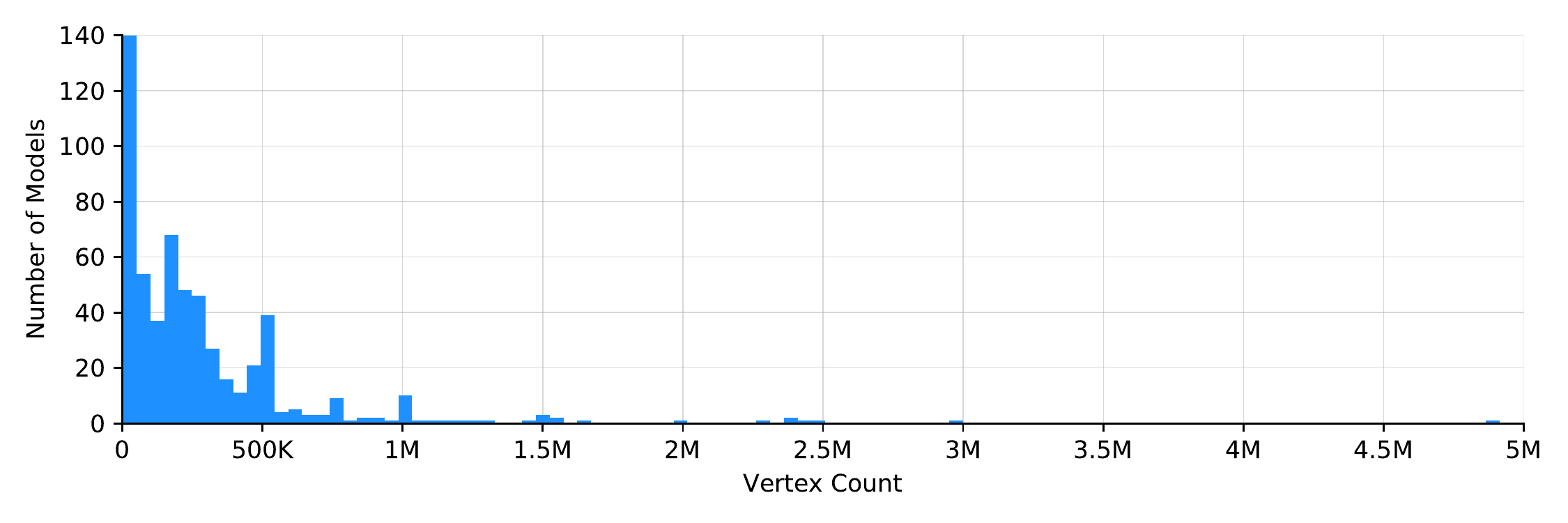}
\caption{Geometric complexity of the 3D models found in our dataset. Models generated from 3D reconstruction technology are typically stored at high resolution to maximize accuracy and fidelity to their real-world counterpart.}
\label{fig:complexity}
\end{figure}

\dataset{} features models of many different sizes, ranging from small models made of only a few thousand triangles to large models composed of millions of faces. The histogram of the geometric complexity of the dataset, expressed as the number of vertices, is reported in figure \ref{fig:complexity}. On average, our models tend to be fairly large compared to those found in similar datasets: as an example, 37\% of the \ds{} models have more than 250K vertices compared to less than 2\% of the models found in Thingi10K \citep{thingi}.
This is expected, as objects are typically reconstructed at high resolution and then optimized for efficient use in downstream applications if needed.

\subsection{Texture Maps and Parametrizations} \label{sec:textures}

\noindent{\bf Texture Size}
Modern photo-reconstruction pipelines are able to generate accurate digital reproductions from hundreds or thousands of high-resolution images.
Consequently, the majority of textures resolutions found in RWTT are ranging from 4K upwards.
Many models feature multiple textures images, including non-squared ones.

\begin{figure}
\centering
\includegraphics[width=\linewidth]{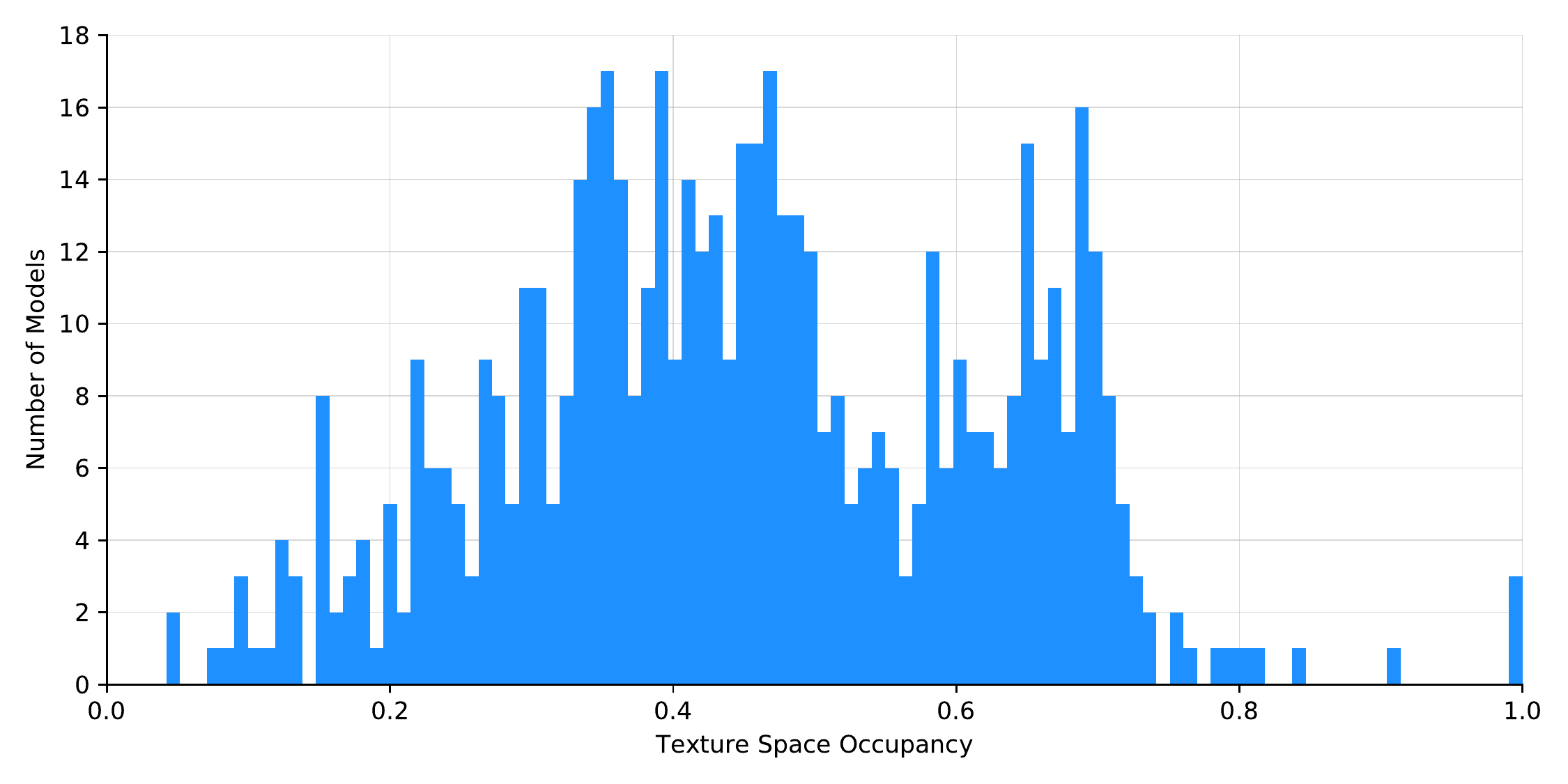}
\caption{Distribution of the \ds{} textures occupancy. Many textures are not efficiently packed, leading to a waste of GPU memory when the models are rendered in real-time applications.}
\label{fig:texocc}
\end{figure}

\noindent{\bf Texture Space Occupancy}
Figure \ref{fig:texocc} plots the histogram of texture occupancy values found in \ds{} and clearly shows how atlas packing is often sub-par, leading to unnecessarily large memory requirements for texture signals sampled at relatively low densities.
This is particularly significant as GPU memory is often a critical bottleneck for graphics applications.

\begin{figure}
\centering
\includegraphics[width=\linewidth]{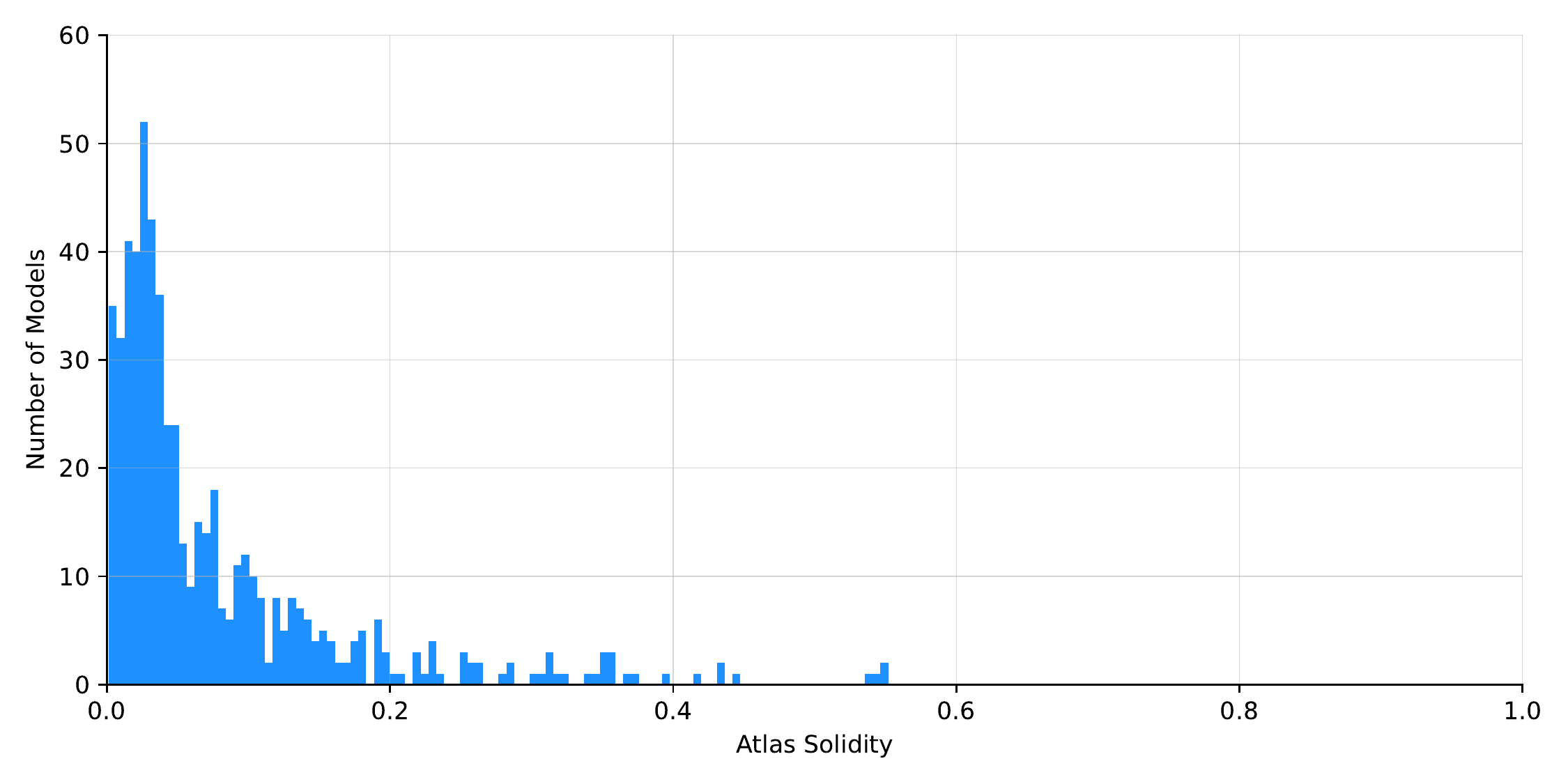}
\caption{Distribution of atlas Solidity for models featured in the \ds{} dataset. Many atlases exhibit high-to-extreme fragmentation: the model with the highest chart count has 555252 charts.}
\label{fig:atlas}
\end{figure}

\noindent{\bf Atlas Solidity}
Figure \ref{fig:atlas} plots the distribution of the atlas solidity across the dataset, and clearly shows a bias towards low solidity values, indicating a very frequent occurrence of texture maps affected by the presence of very large amounts of seams.

The surprising fact, however, is that the degree of fragmentation witnessed in the majority of the models is almost never justified by the topological and geometrical complexity of the models we have collected (see figure \ref{fig:seams}).
We conjecture that this happens because of two distinct causes. First, this happens because the textures are synthesized using large numbers of photographs, with reconstruction tools optimizing for color fidelity and attempting to store as much information as possible from the available images, disregarding the creation of many atlas charts.
Second, this may also happen because of intrinsic limitations in existing automatic UV mapping technology, which we investigate in section \ref{sec:uvtools}.

\begin{figure}
\centering
\includegraphics[width=\linewidth]{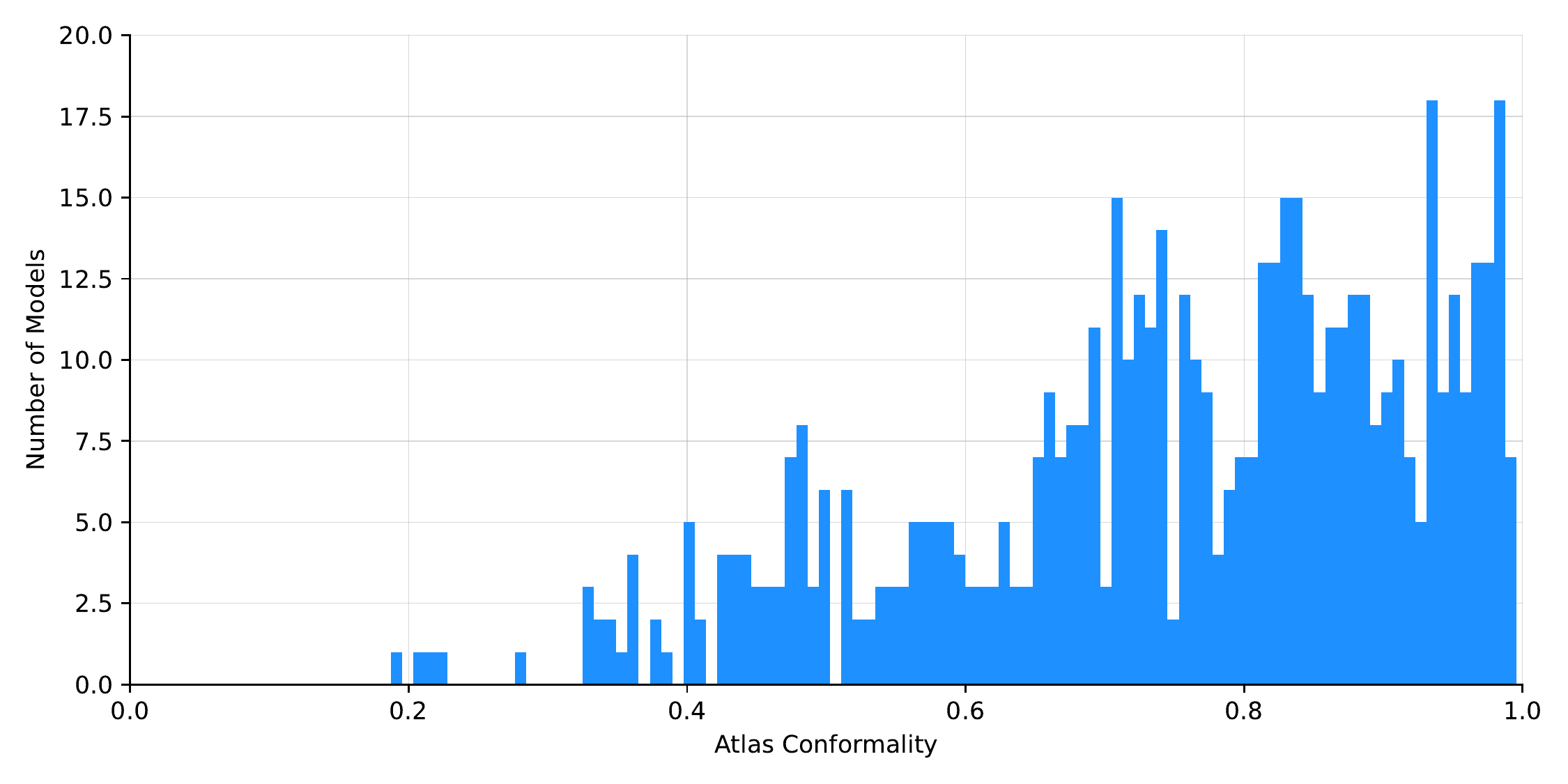}
\caption{Distribution of atlas Conformality, where a lower value implies larger angle deviations. Despite many fine-grain chart decompositions, many parametrizations are still affected by non-negligible conformal distortion. }
\label{fig:distortion}
\end{figure}

\noindent{\bf Conformality and Sampling Frequency}
Figure \ref{fig:distortion} reports the distribution of the average quasi-conformal distortion values for our dataset.
Unsurprisingly, most of the texture maps exhibit good angle preservation; this is to be expected given the high degree of fragmentation affecting many of the atlases.

\begin{figure}
\centering
\includegraphics[width=\linewidth]{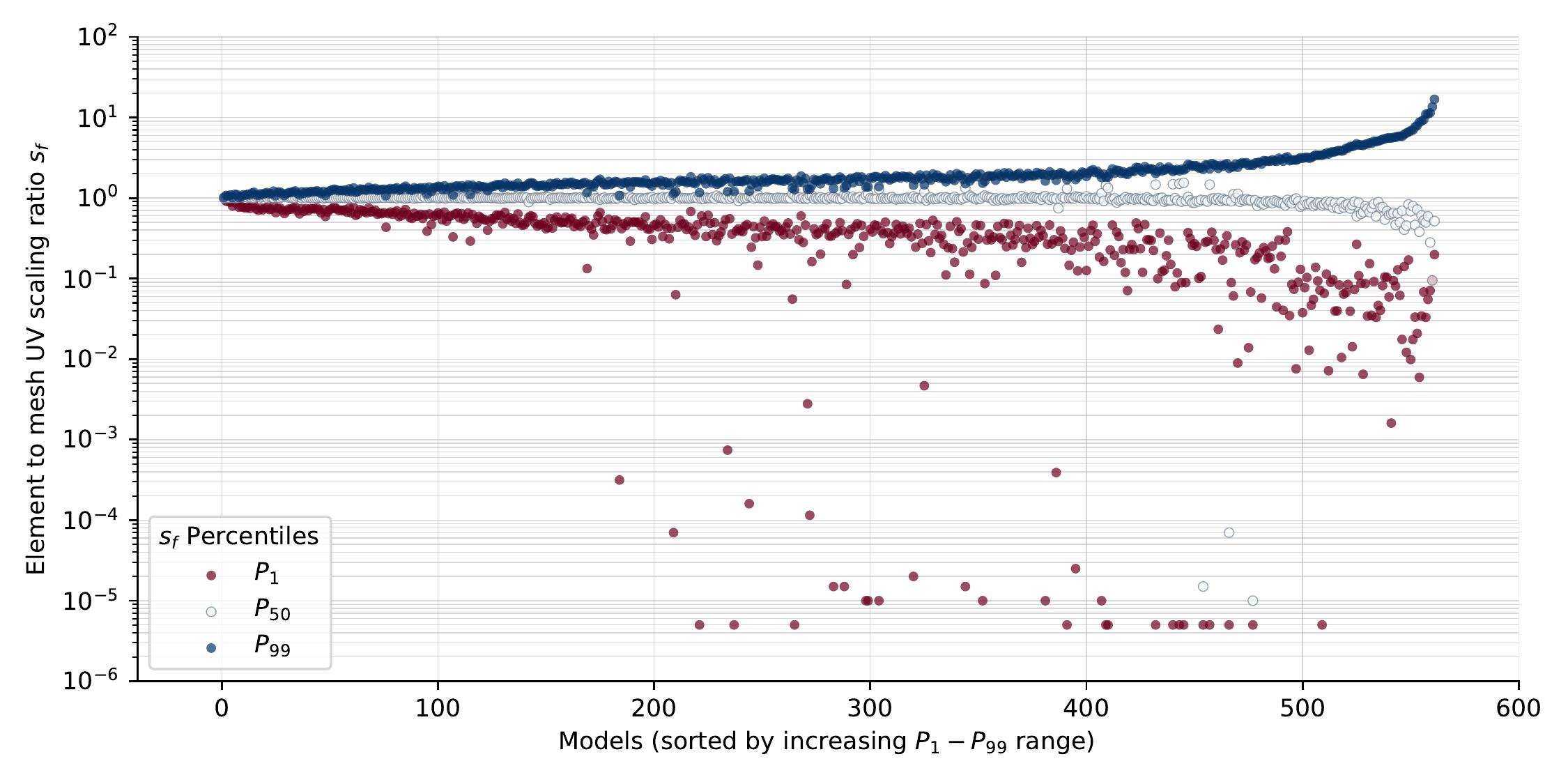}
\caption{Visualization of the UV scaling ranges present in the dataset. We plot the 1st, 50th and 99th percentiles for the models sorted by UV scaling range width, highlighting how a significant portion of the dataset exhibits large variances in the mapping resolution.}
\label{fig:uvscaling}
\end{figure}

A scatter plot of the 1st, 50th and 99th percentile values of the models UV scaling, sorted by range from $P_1$ to $P_{99}$ is reported in figure \ref{fig:uvscaling}, showing how a significant amount of models exhibit a non-negligible spread in the UV scaling values and, consequently, non-uniform texel density across the surface.

Models with less uniform sampling frequency distributions were likely generated from photographs that captured surface details at different scales.

\rev{

\begin{figure}
\centering
\includegraphics[width=\linewidth]{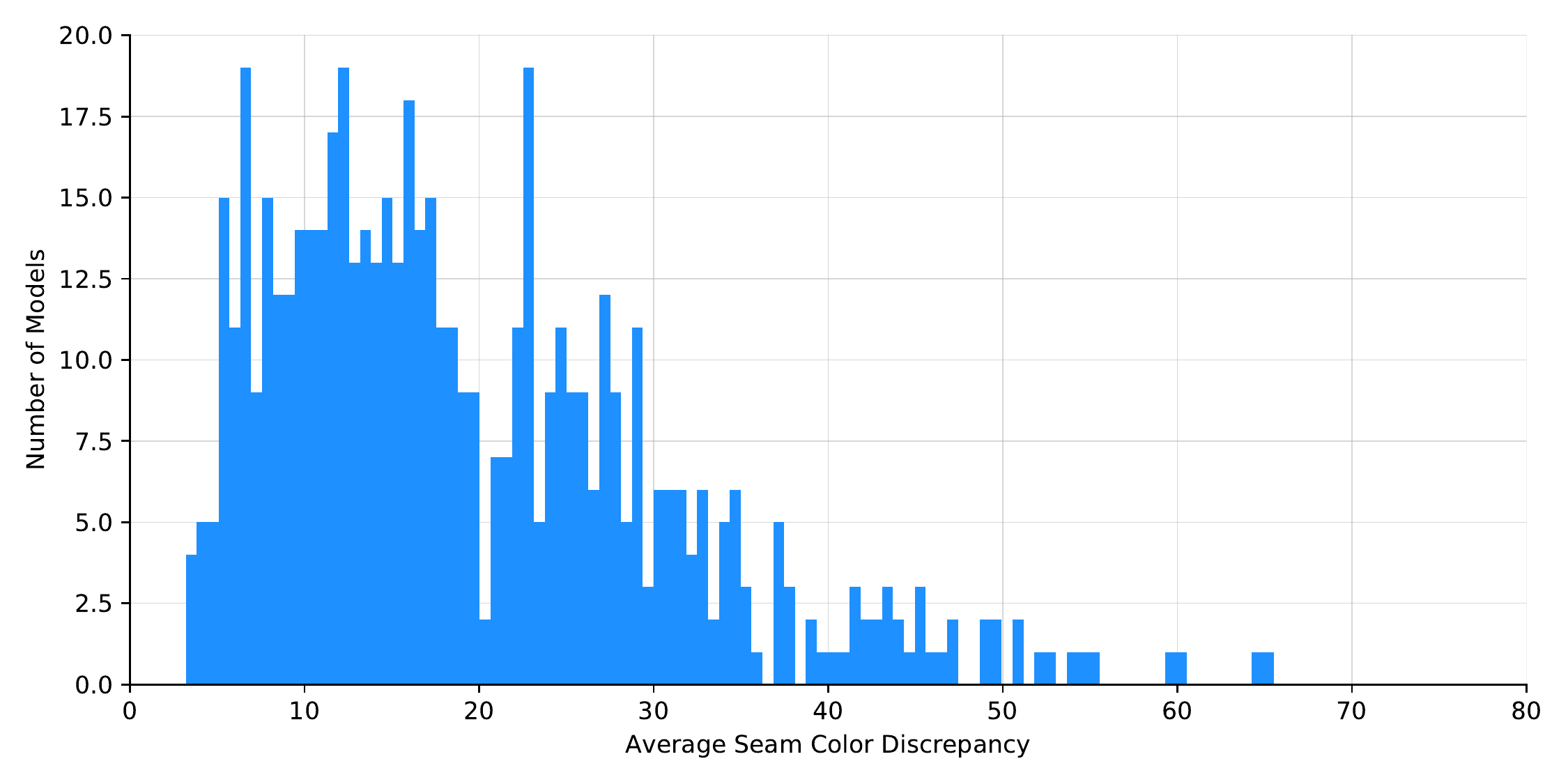}
\caption{Distribution of average seam color discrepancy (measured in RGB space).}
\label{fig:discrepancy}
\end{figure}

\noindent{\bf Seam Color Discrepancy}
The distribution of average color reconstruction discrepancies at texture seams is reported in figure \ref{fig:discrepancy}.
Overall, our dataset offers a fairly heterogeneous representation of color discrepancies in texture seams.
There are two explanations for this diversity.
First, there can be differences in the uniformity of the texture colors whereby models with more varied textures are likely to exhibit more severe artifacts at texture seams compared to models with surface colors that are more uniform.
Second, there can be differences in the way texture seams are handled by the reconstruction tools during the texturing process: some tools may attempt to generate textures that are smoother across seams by optimizing for consistency in the texture data whereas others may prioritize fidelity to the photographic data.

}

\subsection{Defects}

\begin{figure}
\centering
\includegraphics[width=\linewidth]{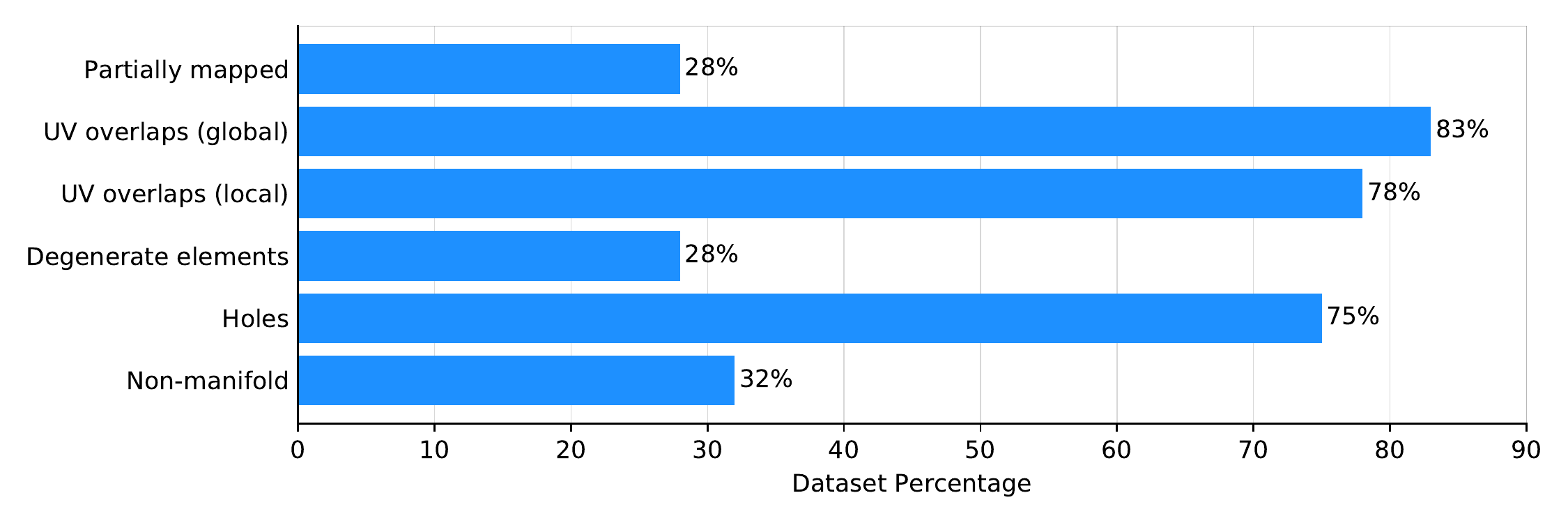}
\caption{Common defects found in \dataset{} models}
\label{fig:defects}
\end{figure}

The ability to deal with arbitrary input meshes without requiring preemptive cleaning is a key property of any robust implementation, whose importance is being increasingly recognized as the emergence of more and more ``real world'' dataset and algorithms specifically developed to deal with them.
\ds{} is a further exemplification of this, and features meshes with all kinds of meshing and geometric defects, as well as defects related to texture mapping such as local and global overlap of UV coordinates, and models that are only partially textured.
An overview of the situation is reported in figure \ref{fig:defects}.

\section{Recomputing UV maps}\label{sec:uvtools}

Given the number of issues these UV maps exhibit, an obvious question is whether or not these results are peculiar to the tools that generated the textures and parametrizations as part of the 3D reconstruction process, or if there is an intrinsic limitation in the current technology regarding automatic UV mapping of complex and high-resolution model.
\rev{Therefore, we have decided to evaluate the performance of existing state of the art UV mapping tools and see if they can produce UV maps of better quality than the ones found in our dataset.}

\rev{
We have performed experiments with commercial, open-source and academic automatic UV mapping solutions.
Maya \citep{maya2019} is a commercial 3D modeling software that provides a simple automatic UV unwrapping procedure by projecting 3D faces to a fixed set of orthogonal planes.
Blender \citep{blender} also has automatic projection-based UV unwrapping functionality, but the set of projection planes is not fixed: instead, charts are computed by grouping together faces with similar normal vectors, and the projection plane of each chart is orthogonal to the average normal.
Finally, we also experimented with OptCuts \citep{Li:2018:OptCuts}, which is a recent academic work that automatically computes UV maps by alternating steps in which the distortion of the mapping is reduced, and steps in which cuts are added to the atlas charts to further reduce distortion.
}

For our experiments, we decided to set a time limit on the overall running time of 20 minutes. Since these tools cannot conveniently produce multiple textures (rather, UV channels) we have restricted the experiment only to models with single textures.
We also restricted the experiments to models comprised of a unique connected component as \rev{the available implementation of OptCut fails to initialize multi-component models.}
To meet the topological requirements of OptCuts (the input must be two-manifold), we implemented a simple preliminary cleaning step that removes non-manifold vertices and edges, unreferenced vertices, and degenerate geometry. Moreover, since OptCuts is an iterative algorithm we saved the result obtained after 20 minutes of running time, rather than simply interrupting the task.

\begin{figure}
\centering
\includegraphics[width=0.4\linewidth]{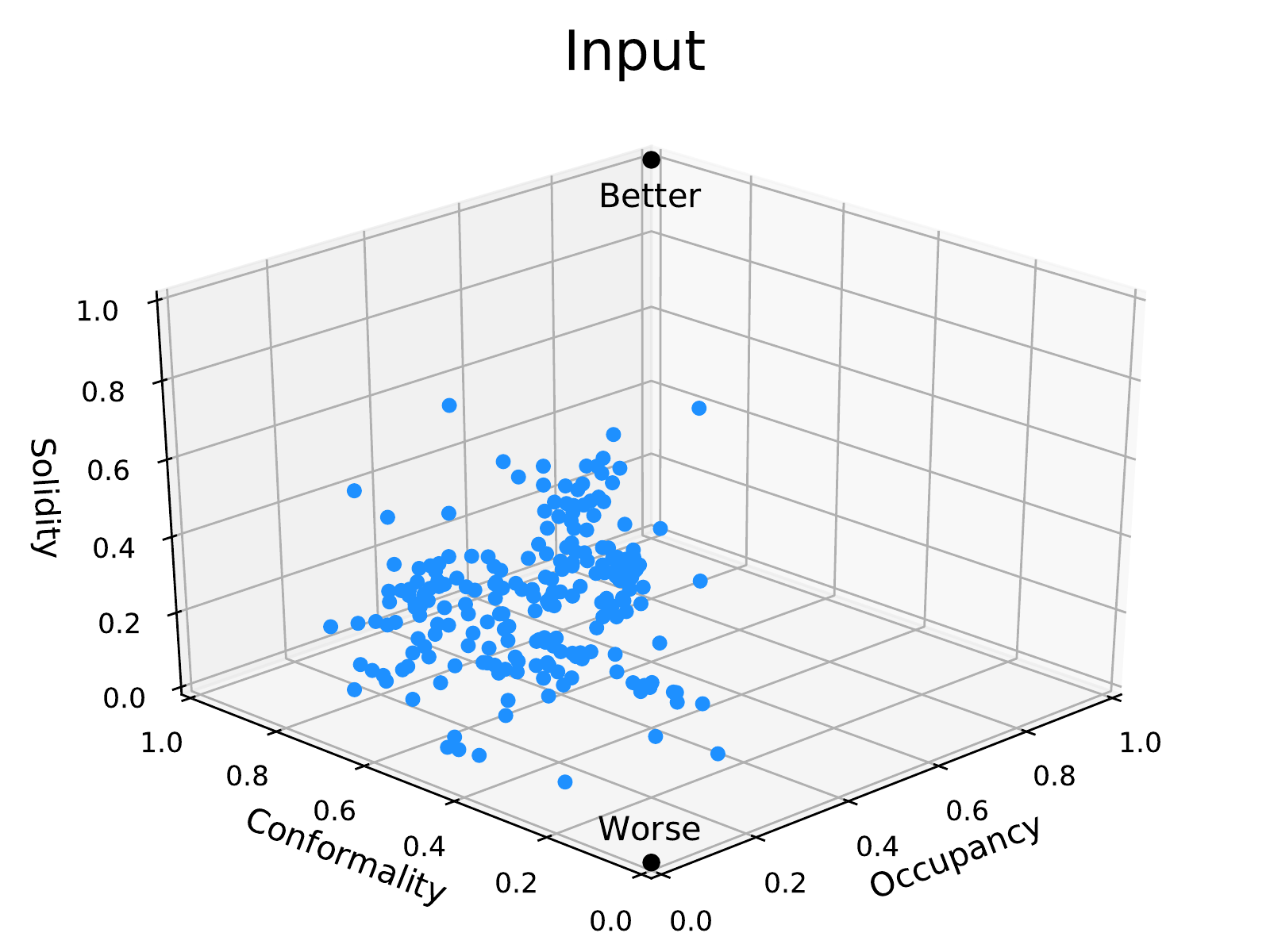}
\includegraphics[width=0.4\linewidth]{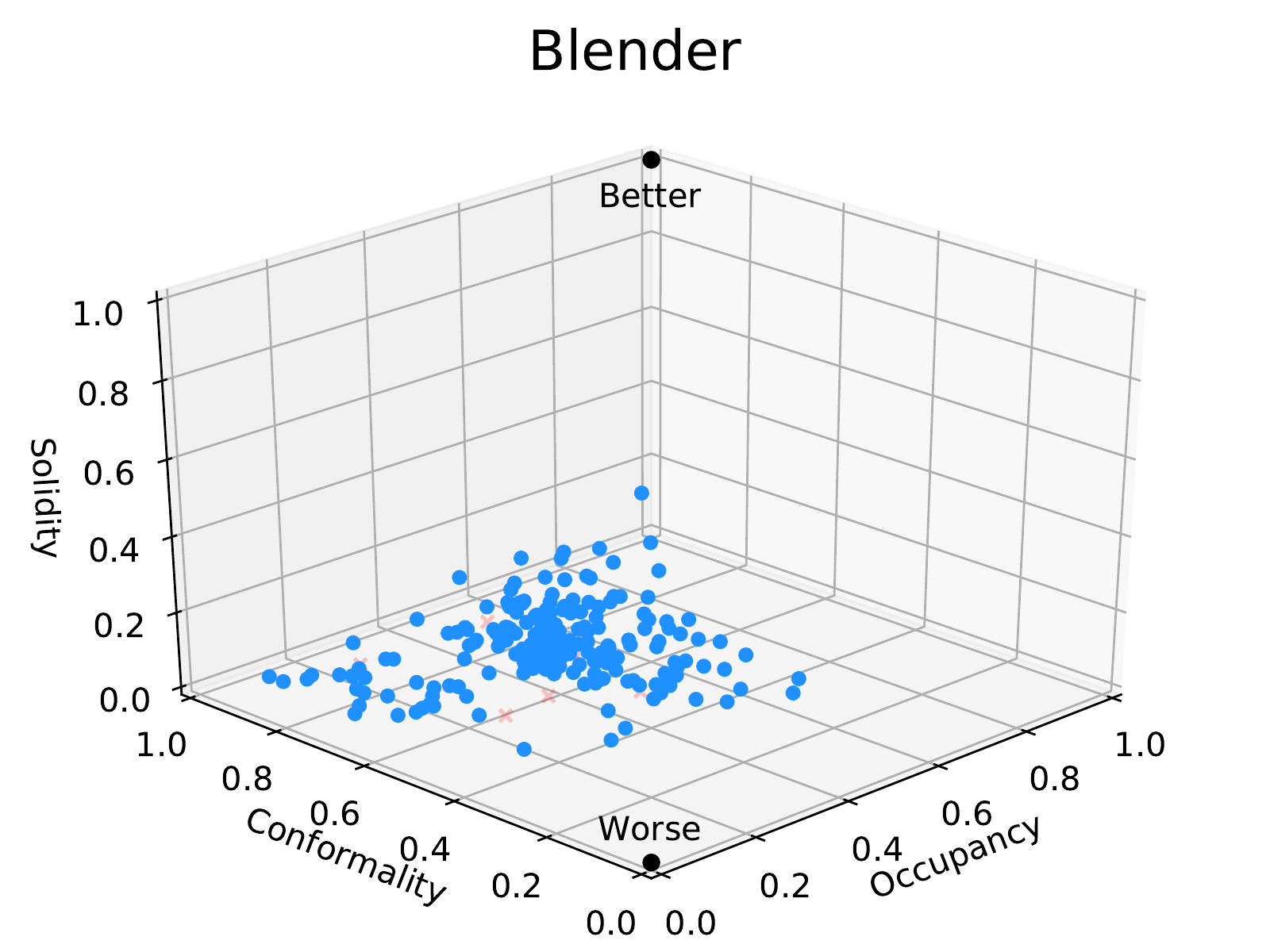} \\
\includegraphics[width=0.4\linewidth]{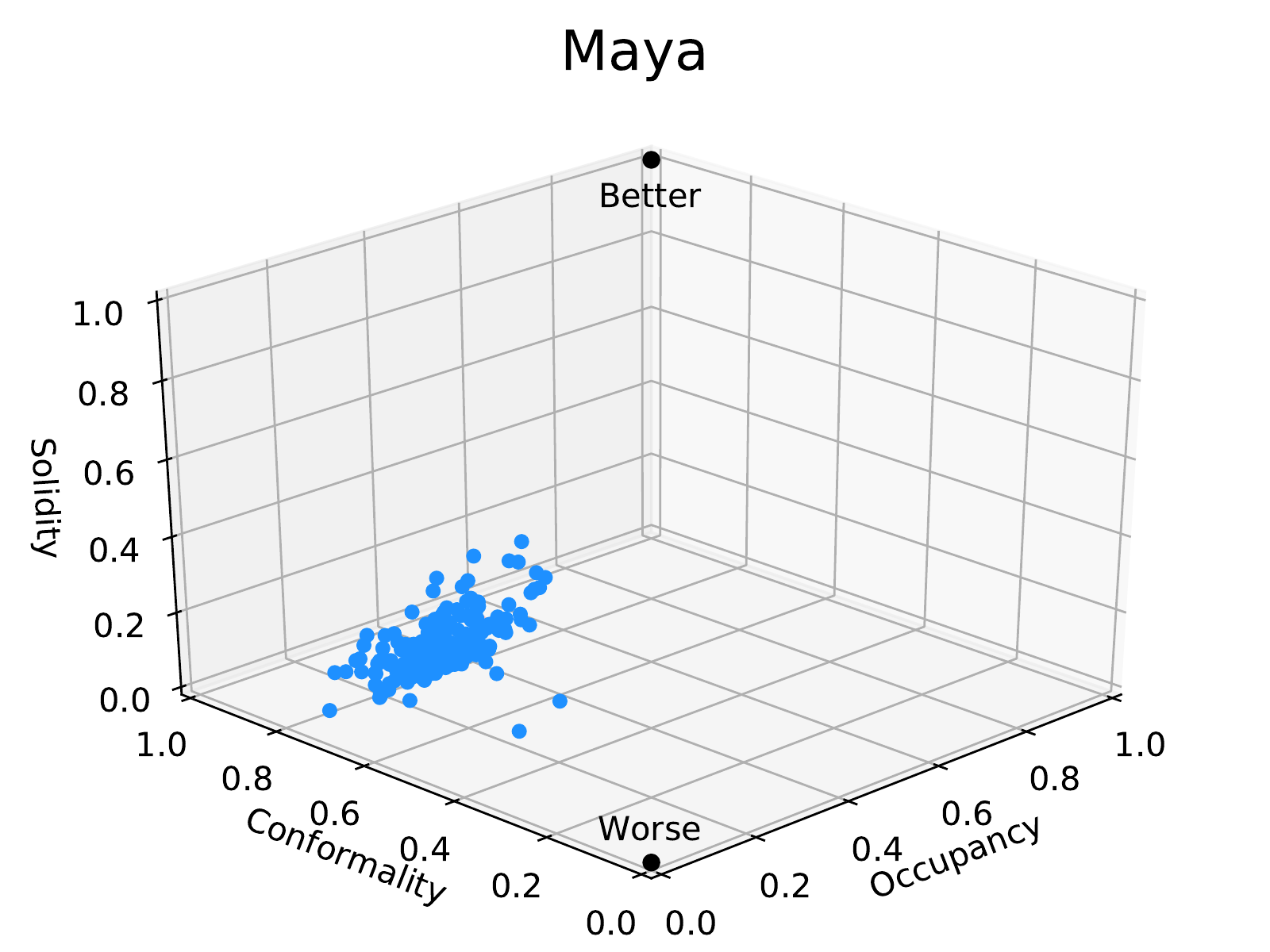}
\includegraphics[width=0.4\linewidth]{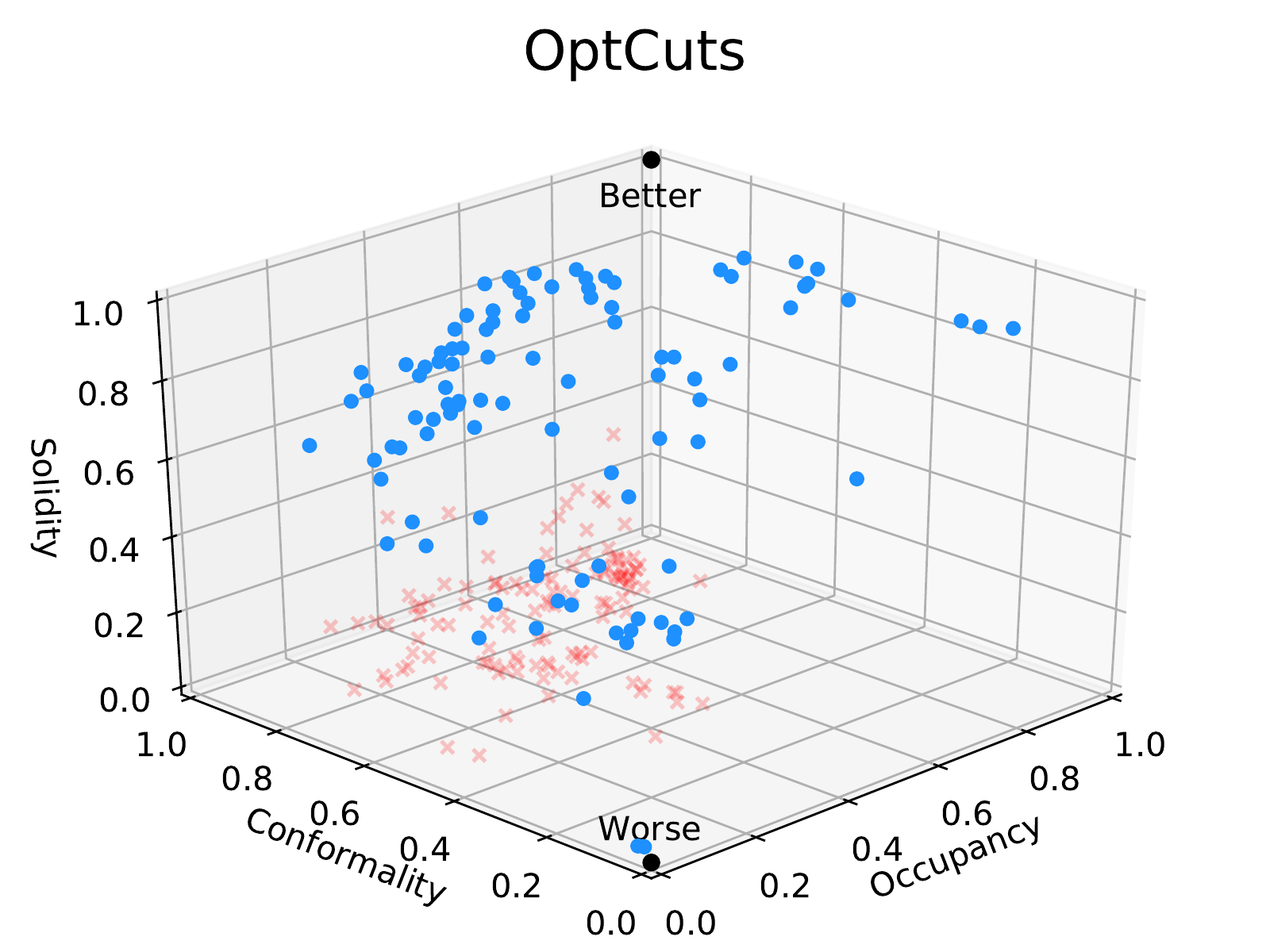}
\caption{Performance of existing UV mapping tools on models generated with photo-reconstruction technology.}
\label{fig:uvtools}
\end{figure}

The results of this experiment are reported in figure \ref{fig:uvtools} as a 3D scatter plot of the texture occupancy, mapping conformality, and atlas solidity of the input data and the outputs produced by the three implementations used in the experiment, with failure cases marked in red.
Ideally, the scatter plots of the outputs should move towards the top corner at coordinates (1, 1, 1).
\rev{Blender performs worse in terms of atlas solidity compared to the input UV maps, without managing to improve under any of the remaining two metrics.
Maya also generates heavily fragmented output atlases, but succeeds in  significantly reduce the conformal distortion of the UV maps; note however that} the results are also affected by packing issues as shown by the poor performance in terms of texture occupancy.
OptCuts shows the best results despite being often interrupted before completing the atlas optimization, but it is affected by a very high failure rate (54\%). OptCuts failures are either due to numerical issues that prevent the correct initialization of the UV configuration, or inability to reduce the surface genus before starting. Finally, it is worth mentioning that, of the three implementations we tested, OptCuts is the only one that guarantees the bijectivity of the mapping; however, it is a much less efficient, practical \rev{and reliable} solution.

\rev{Overall, this experimental evaluation shows that automatically computing UV maps for this category of 3D models is intrinsically challenging regardless of the method employed, and current technology is unable to provide satisfactory results in most cases.}

\section{The \texmetro{} Tool}\label{sec:texmetro}

Alongside the dataset, we also release an open-source command line tool (\texmetro{}) that can be used to compute all the measures
described in Sec.~\ref{sec:measures} on arbitrary user provided photo-reconstruction models.
This tool can be used to assess and compare existing or future alternative photo-reconstruction tools (e.g. starting from the same set of images),
as well as to assess the efficacy of existing or future methods aimed to improve the quality of 3D models.

The UV-map, geometry, and connectivity of the an input models are analyzed.
The statistics in full are collected in a JSON file while only a relevant subset of them is displayed on the standard output, consisting of salient information about the mesh geometry, its parameterization and texture data.
A description of the JSON file structure can be found in the documentation from the \texmetro{} repository.

A few of the measures are dependent on the texture resolution (in texels). To compute these, \texmetro{} takes as input the actual texture image, or, simply, its dimensions, provided as command-line arguments.

The \texmetro{} tool is open-source and is implemented using C++, Qt (for interfacing), and OpenGL (to accelerate texture overlap detecting, which is implemented by rasterization over off-screen buffers).

\section{Conclusions} \label{sec:conclusions}

In this work, we introduce a dataset of textured 3D models generated with automatic software, carefully selected and gathered from publicly available sources, and enriched with relevant measures, including new ones tailored to highlight the defects of this increasingly important (but often neglected) class of 3D models; we also provide an open-source tool to collect measures and statistics on the geometry, parametrization and texture maps of user-provided models.

The analysis of our dataset reveals interesting traits common to these models.
In particular, it is extremely common to find heavily fragmented texture atlases; this could be attributed to the fact that these parametrizations are generated automatically and tailored to the available pictures.
Moreover, experimental evaluation of current state of the art automatic UV mapping tools, both commercial and academic, suggests that the issues that affect the UV maps of our dataset are not peculiar to the implementations of the software used to reconstruct the mesh; rather, this seems to be a limitation of current automatic UV mapping tools and algorithms that either fail to produce better results or are not practical and robust enough to reliably undertake the task.

Our aspiration is that this dataset and tool will help researchers and developers in their efforts to improve the quality of photo-reconstruction algorithms and implementations, as well as new tools targeted at post-processing the resulting models.

\section*{Acknowledgements}
 This research was partially funded by the Italian PRIN project {\em DSurf} (grant no. 2015B8TRFM) and by the EU project {\em PARTHENOS: Pooling Activities, Resources and Tools for Heritage E-Research Networking, Optimization and Synergies} (grant no. 2015-2019 H2020-INFRADEV-1-2014).

\bibliography{biblio}

\end{document}